\documentclass[amsmath,amssymb,pra,reprint]{revtex4-1}

\usepackage{graphicx}
\usepackage{dcolumn}

\newcommand{\ket}[1]{|{#1}\rangle}
\newcommand{\bra}[1]{\langle{#1}|}

\newcommand{\ketbra}[2]{\ket{#1}\bra{#2}}

\newcommand{\cL}{{\mathcal L}}

\newcommand{\cQ}{{\mathcal Q}}
\newcommand{\cK}{{\mathcal K}}
\newcommand{\cD}{{\mathcal D}}
\newcommand{\cP}{{\mathcal P}}
\newcommand{\Tr}{{\rm Tr}}

\renewcommand{\Re}{{\rm Re}}
\newcommand{\damp}[1]{{\mathcal D}\!\!\left[#1\right]}

\begin{document}


\title{Optomechanical laser cooling with mechanical modulations}

\author{Marc Bienert}%
\affiliation{Theoretische Physik,
  Universit\"at des Saarlandes, D-66123 Saarbr\"ucken, Germany }%

\author{Pablo Barberis-Blostein}%
\affiliation{Instituto de
  Investigaciones en Matem\'aticas Aplicadas y en Sistemas,
  Universidad Nacional Autonoma de M\'exico,  Circuito Escolar s/n
  Ciudad Universitaria
  M\'exico, D.F. }


\date{\today}

\begin{abstract}
 We theoretically study the laser cooling of cavity optomechanics when the mechanical resonance frequency and damping depend on time. In the regime of weak optomechanical coupling we extend the theory of laser cooling using an adiabatic approximation. We discuss the modifications of the cooling dynamics and compare it with numerical simulations in a wide range of modulation frequencies.
\end{abstract}

\maketitle


\section{Introduction}
\label{sec:Intro}

Quantum cavity optomechanics~\cite{reviewOM} deals with the physics of
a mechanical element coupled to the light field of an optical
resonator by radiation forces in the quantum regime. In the simplest
and most commonly used model a single mechanical harmonic oscillator
is coupled to a single mode of an optical cavity by radiation pressure
interaction being proportional to the light intensity. The prototype
of an optomechanical setup is a Fabry-Perot cavity with pendular
end-mirror, whose microfabricated
realization~\cite{om:groeblacher2009}, but also other
implementations~\cite{om:verhagen2012,om:eichenfeld2009,om:korppi2013,flowers:2012},
have been developed in the laboratories worldwide. The optomechanical
coupling allows for optical control of the mechanical object,
manifesting itself in strongly modified mechanical properties what
ultimately can lead to laser
cooling~\cite{om:mancini1994,om:wilsonrae2007,om:marquardt2007}
towards the mechanical ground state, as has been demonstrated in a
certain setup~\cite{om:chan2011}, marking a requisite milestone on the way to quantum
applications.

For typical realizations, the optomechanical coupling is clearly
smaller than the mechanical frequency. In such a regime, laser cooling
can be considered as a consequence of Raman scattering of photons
with Stokes- and anti-Stokes events, where a single
vibrational quantum is deposited or taken away with the scattered
photon. These processes play the central role in the cooling dynamics, similar as in the Lamb-Dicke regime of
laser cooling of atoms. When Anti-Stokes scattering prevails, the
mechanical system is cooled. The balance between Stokes- and
Anti-Stokes scattering can be adjusted by the laser parameters and for resolved
sideband cooling, when the cavity's linewidth is smaller than the
mechanical frequency $\nu$, the pump laser optimally has to be detuned
to the red side of the cavity resonance by $\nu$.

The optomechanical coupling strength is typically even so weak, that
cooling with single photons turns out to be much too slow to overcome the
rethermalization rate due to the coupling with the mechanical element's
environment at cryogenic or even room temperature. To overcome this
obstacle, in many setups a strong pump laser is used to effectively
boost the optomechanical interaction. Beyond that, more sophisticated
cooling schemes are based on pulsed pump schemes exploiting
interference effects~\cite{liao:2011,machnes:2012}, or dynamically
controlled cavity dissipation~\cite{liu:2013} in order to improve the
cooling efficiency. Short optical pulses can also be used for state
preparation and reconstruction~\cite{Vanner:2011}, and modulated driving
allows for the generation of squeezed quantum states~\cite{mari:2009}.
The modulation of the mechanical frequency was investigated in
Ref.~\cite{farace:2012} with the result that squeezing and
entanglement are enhanced in an resonant manner when the modulation
frequency is twice the mechanical frequency, while the mean
vibrational occupation number is simultaneously increased.

In this work, we focus on the behavior of laser cooling when the
mechanical frequency and the mechanical damping is varied in time.
This investigation was motivated by the recent experimental
work~\cite{om:joeckel2011} using dielectric membranes oscillatingly
mounted inside a Fabry-Perot cavity. By local heating of the membrane
with the help of a laser beam, both the resonance frequency of a
vibrational mode and the linewidth can be altered in a controlled way.
We extend the theory of optomechanical cooling to include
modulations of the mechanical frequency in an adiabatic way and
compare the predictions with numerical results. Moreover, we discuss
how the modulated damping rates influences the cooling behaviour. We
find that in the resolved sideband limit and for strong periodic
modulations, additional cooling resonances appear. For weaker
modulations or larger cavity linewidths these resonances overlap and
with the help of different pulse shapes, the form of the resulting
resonances can be influenced. Moreover, we systematically scan a large
range of modulation parameters and discuss the cooling behavior in
the various regimes.

This article is set up as follows: We first provide a theoretical
description of the model in Sec.~\ref{sec:model} by presenting the
master equation of the open quantum system and the linearized model
used for numerical calculations. In Sec.~\ref{sec:cooling} we extend
the theory of optomechanical cooling by including the temporal changes of the
mechanical frequency into the description using an adiabatic
approximation. Moreover, we point out the changes due
to the modulation and compare them with numerical calculations. In
Sec.~\ref{sec:tierra}, we then focus on a larger range of modulation
frequencies and damping rates of the mechanical oscillator. We discuss
the numerical findings and connect them to the theory
of modulated cooling. Finally, in Sec.~\ref{sec:conclusions}, we
summarize and draw the conclusions.

\section{Modulated optomechanical setup}
\label{sec:model}

\begin{figure}
 \begin{center}
  \includegraphics[width=6.5cm]{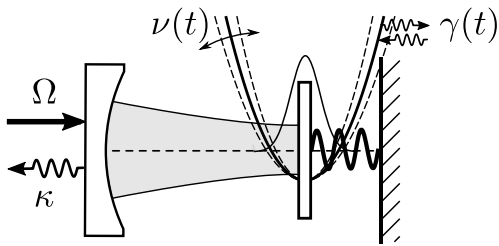}
 \end{center}
 \caption{\label{fig:model}An optomechanical setup, consisting of an optical cavity coupled to a harmonically supported mechanical element by radiation forces, experiences modulations of the mechanical properties, {\it i.e.} periodically changes in the mechanical frequency $\nu(t)$ and damping rate $\gamma(t)$. The cavity is pumped by an external laser at rate $\Omega$ while photons leak out of the cavity at rate $2\kappa$. For certain parameters, the light scattering leads to cooling of the mechanical motion, which is influenced by the mechanical modulation.}
\end{figure}

\subsection{Model} We consider an optomechanical setup consisting of a
single mechanical oscillator of effective mass $M$ coupled by radiation forces to a single
driven mode of an optical cavity with frequency $\omega_\text{cav}$, see Fig.~\ref{fig:model}.
The Hamiltonian for such a setup in the frame rotating with the
frequency $\omega_\text{P}$ of the pump laser reads
\begin{align}
 H(t) = H_\text{cav} + H_\text{mec}(t) + H_\text{rad} + W
 \label{eq:H}
\end{align} whereby
\begin{align}
 H_\text{cav} &= -\hbar\delta a^\dagger a\\ H_\text{mec}(t) &=
 \frac{p^2}{2M}+\frac12 M \nu(t)^2 x^2 
\end{align} are the unperturbed Hamiltonians of the cavity mode with
detuning $\delta=\omega_\text{pump}-\omega_\text{cav}$ and the
mechanical oscillator with time-dependent frequency $\nu(t)$,
respectively. The annihilation operator $a$ of a single photon in the
cavity obey $[a,a^\dagger]=1$. The mechanical
oscillator's position and momentum operators $x$ and $p$ are connected
to the time-independent annihilation operator by
\begin{align}
 b_0 = \frac{1}{2\xi_0} x + i \frac{\xi_0}{\hbar} p
\end{align} where we introduced the harmonic oscillator's length scale
\begin{align}
 \xi_0 = \sqrt{\frac{\hbar}{2 M \nu_0}},
\end{align} 
with the time-averaged oscillator frequency $\nu_0 = \frac1{T}\int_0^T \nu(t) dt$ over one period $T$ of the modulation. The optical and
mechanical degree of freedom is coupled by radiation forces of the form
\begin{align}
 H_{\rm rad} &=-\hbar g a^\dagger a \, x\nonumber\\
&= -\hbar\chi_0 a^\dagger a[b_0+b_0^\dagger].
\end{align} 
Here, $\hbar\chi_0/\xi_0=\hbar g$ can be interpreted as the time-independent radiation force
a single photon exerts on the unmodulated mechanical oscillator. In a Fabry-Perot
setup like in Fig.~\ref{fig:model}, $\chi_0=\omega_\text{C}\,\xi_0/L$, where
$L$ denotes the static cavity length. For the membrane in the middle setup, an additional factor including the membranes electric field reflectivity and its position within the unperturbed mode function of the cavity has to be included~\cite{om:Jayich2008,om:wallquist2010}.
The pump of the cavity is taken into account by
\begin{align}
 W = \hbar \frac{\Omega}{2}(a+a^\dagger).
\end{align}

With the Hamiltonian $H$, Eq.~\eqref{eq:H}, we can write down the master equation
\begin{align}
 \frac{\partial \rho}{\partial t}
  &= \frac{1}{i\hbar}[H(t),\rho] + \cL_\kappa\rho + \cL_\gamma(t)\rho
  \label{eq:liou}
\end{align} where
\begin{align}
 \cL_\kappa\rho &= \kappa \cD[a]\rho\\ 
\end{align} 
describe cavity losses with decay rate $\kappa$. The damping of the oscillator with a time dependent damping rate $\gamma(t)$ we approximately describe by
\begin{align}
  \cL_\gamma(t)\rho &=
 \frac{\gamma(t)}2 (\bar m+1) \cD[b_0]\rho + \frac{\gamma(t)}2 \bar m
 \cD[b^\dagger_0]\rho\label{eq:Lgamma}
\end{align}
using the time-independent operators $b_0$, $b^\dagger_0$, leading to reasonable description for moderate modulations stengths~\cite{hanggi:1997}. The temperature $T$ of the environment thermalizes the oscillator towards a mean vibrational quantum
number $\bar m=[\exp(\hbar
\nu_0/k_{\rm B} T)-1]^{-1}$ in the unmodulated case.
We have used the short notation $\cD[X]\rho=2 X\rho
X^\dagger-\{X^\dagger X,\rho\}$ for a Lindblad-form term.

\subsection{Displaced frame}

In order to eliminate the pump, we unitarily transform $\rho = U\rho'
U^\dagger$ with the help of
\begin{align}
 U=D_a(\alpha(t))D_{b_0}(\beta(t))
\end{align} consisting of the displacement operators
$D_a(\alpha)=\exp[\alpha a^\dagger - \alpha^\ast a]$ for the cavity
and the $D_{b_0}(\beta)=\exp[\beta b_0^\dagger - \beta^\ast b_0]$ for the mechanical oscillator. In the
displaced picture, the master equation reads (primes at $\rho$ omitted)
\begin{align}
  \frac{\partial \rho}{\partial t} &= \cL\rho\nonumber\\ &=
  \frac{1}{i\hbar}[H_\text{cav}'+H_\text{mec}(t)+H_\text{rad}',\rho] +
  \cL_\kappa\rho + \cL_\gamma\rho,
  \label{eq:me}
\end{align} if the parameters $\alpha(t)$ and $\beta(t)$ are chosen
such that they fulfill the differential equations
\begin{subequations}
\label{eq:dsysab}
\begin{align}
 \dot\alpha(t) &=
 \left\{i[\delta+2\chi_0\Re\beta(t)]-\kappa\right\}\alpha(t)-i\frac{\Omega}{2},\\
 \dot\beta(t) &=
 -\left\{\frac{\gamma(t)}{2}+i\nu_+(t)\right\}\beta(t)+i \beta^\ast\nu_-(t)+i\chi_0|\alpha(t)|^2,
\end{align}
\end{subequations} together with their complex conjugated
counterparts and $\nu_\pm(t) = [\nu_0\pm\nu^2(t)/\nu_0]/2$. Then, the transformed Hamiltonians
take on the form
\begin{align}
 H_\text{cav}'&=-\hbar\delta'(t)a^\dagger a\\
 H_\text{rad}'&=-\hbar\chi_0 \left[a^\dagger a+\alpha(t)a^\dagger
 +\alpha^\ast(t) a\right](b_0+b_0^\dagger)\label{eq:Hrad}
\end{align} with $\delta'=\delta +2\chi_0\Re\beta(t)$. We assume that
the cavity is strongly driven, such that $|\alpha|\gg 1$, boosting in
that way the linear optomechanical coupling.

\subsection{Linearized interaction}
\label{sec:linint}

If the cavity is sufficiently strongly pumped, the mean field $\alpha$
becomes much larger than the field fluctuations and the term
proportional to $a^\dagger a$ in the interaction
Hamiltonian~\eqref{eq:Hrad} can be omitted~\cite{om:wilsonrae2008}.
The master equation~\eqref{eq:me} can then brought into a quadratic form~\cite{wallquist2010}
\begin{align}
 \frac{\partial\rho}{\partial t}=\frac{1}{i\hbar}\left[\vec R^\text{T} \hat H \vec R,\rho\right]+\sum_{k=1}^3\frac{\gamma_k}{2}\damp{\vec L_k\vec R}\rho,
 \label{eq:meqf}
\end{align}
where the vector operator $\vec R=(x_\text{c}, p_\text{c}, x_\text{m},
p_\text{m})^{\rm T}$ contains the dimensionless position and momentum
operators $x_\text{m}=(b_0+b_0^\dagger)/\sqrt{2}$ and
$p_\text{m}=(b_0-b_0^\dagger)/\sqrt{2}i$ and the corresponding expression
for $x_\text{c}$, $p_\text{c}$ associated with the cavity field mode.
The expectation values of position and momentum and the covariances of
the state $\rho$ are contained in the expressions
\begin{align}
 \vec r &= \langle \vec R\rangle\\
 (\hat C)_{ij} &= \frac12 \langle R_i R_j + R_j R_i\rangle -\langle R_i\rangle\langle R_j\rangle.
\end{align}
They fulfill the dynamical equations
\begin{align}
 \frac{\partial}{\partial t}{\vec  r}(t) &= \hat H_\text{eff}\vec r(t),\label{eq:dynr}\\
 \frac{\partial}{\partial t} \hat C(t) &=\hat H_\text{eff} \hat C(t) + \hat C(t) \hat H_\text{eff}^\text{T}+\hat J\label{eq:dynC}
\end{align}
following from Eq.~\eqref{eq:meqf}. Here, $\hat
H_\text{eff}=2\hat\sigma [\hat H + \text{Im}\hat\Gamma]$ and $\hat J =
2\hat\sigma [\text{Re}\hat\Gamma]\hat\sigma^\text{T}$, whereby the
matrix $\hat \Gamma$ incorporates the dissipative dynamics and is
defined in terms of the quantities $\vec L_k$ in App.~\ref{app:linme}.
Moreover, we introduced the symplectic form
$\hat\sigma_{ij}=[R_i,R_j]/i$. Equations~\eqref{eq:dynr}
and~\eqref{eq:dynC} together with the solution of
Eq.~\eqref{eq:dsysab} are used for a numerical propagation of an
initial thermal (Gaussian) state with mean vibrational number $\bar m$
towards the quasi-stationary solution in a stable parameter regime. The
mean phonon number
\begin{align}
 \langle m\rangle(t) = \frac12\left[C_{33}(t)+C_{44}(t)-1\right].\label{eq:mfromC}
\end{align}
can then be extracted from the propagated covariance matrix $\hat C$.

\section{Cooling the mechanical motion}
\label{sec:cooling}

In this section we study the cooling of the mechanical object with modulated frequency and damping in different parameter regimes. Throughout this work we consider a weak optomechanical interaction, such that $|\alpha|\chi_0/\nu\ll 1$. Cooling and heating is then achieved by light scattering at the cavity into the Anti-Stokes and Stokes components, which are associated with the annihilation and creation of a single vibrational quantum. If the cooling rate $A_-$, {\it i.e.} the rate of Anti-Stokes scattering is larger than the heating rate $A_+$ belonging to Stokes scattered light, the mechanical oscillator converges to the stationary state of laser cooling. In the following we develop a theoretical description of the stationary state of cooling for a modulated oscillator.

\subsection{Perturbation theory and adiabatic approximation}

In order to get insight into the dynamics when the mechanical
element's properties periodically depend on time, we derive an
effective master equation for the determination of the stationary
state. The applied procedure relies on the weak mechanical coupling,
expressed by $\eta= |\alpha|\chi_0/\nu\ll 1$. As a consequence, the
dynamics splits into a hierarchy of time scales~\cite{om:wilsonrae2008}: A fast time scale
given by the free evolution of cavity and oscillator, and a slower
time scale on which the optomechanical interaction takes place. The time scale of the modulation we also assume here to be fast, {\it i.e.} of the order of the mechanical frequency $\nu$. We
expand the Liouvillian $\cL$ from Eq.~\eqref{eq:me} according to
\begin{align}
 \cL = \cL_0 + \cL_1.
 \label{eq:cl01}
\end{align} into different orders in the small parameter $\chi_0/\nu$
with
\begin{align}
 \cL_0(t)\rho &= \cL_\text{c}+\cL_\text{m}(t)\nonumber\\ &=
 \left(\frac{1}{i\hbar}[H_\text{cav},\rho] + \cL_\kappa\rho\right) +
 \frac{1}{i\hbar}[H_\text{mec}(t),\rho],\\ \cL_1\rho &=
 \frac{1}{i\hbar}[H_\text{rad}',\rho],\label{eq:L0}
\end{align} where we neglected the small correction of the cavity
detuning and omitted the damping of the high-$Q$ oscillator
momentarily.

In lowest order in $\eta$, the optical and mechanical degree of freedom are decoupled, {\it i.e.} both degrees of freedom evolve independently. The Liouvillian depends explicitly on time due to the modulation of the mechanical frequency. In App.~\ref{app:specdecomp} we introduce its spectral decomposition 
\begin{align}
 \cL_0(t) = \sum_\lambda \lambda(t) \cP^{\lambda}_0(b_0,b_0^\dagger),
 \label{eq:adiabapp}
\end{align}
with the eigenvalues
\begin{align}
 \lambda_0(t) = \lambda_\text{c}+\lambda_\text{m}(t)
\end{align}
being a sum of the eigenvalues of $\cL_\text{c}$ and $\cL_\text{m}$, whereby $\lambda_\text{m}(t) = i k \nu(t)$ with integer $k$. In Eq.~\eqref{eq:adiabapp} we neglected the time-dependency of the projectors $\cP^{\lambda}_0$ and use the projectors of the unmodulated oscillator: This is an adiabatic approximation being valid as long as transitions between subspaces belonging to different $\lambda$ can be neglected during the time evolution. One can estimate the condition $(\langle m\rangle+1)\frac{\dot \nu}{4\nu^2}\ll 1$ for this approximation to be valid by considering the overlap between energy eigenstates and their derivative with respect to time. 

In lowest order of perturbation theory, all projectors on mechanical energy eigenstates are quasi-stationary states, since their eigenvalues $\lambda_\text{m}$ vanish. In higher order in the optomechanical interaction, this degeneracy is lifted, thereby singling out the unique asymptotic state of laser cooling. 

The cavity and oscillator degrees of freedom are coupled in first
order by the Liouvillian $\cL_1$. In $\cL_1$ we can replace
$\alpha(t)$ with the solution of Eqs.~\eqref{eq:dsysab} in lowest
order in $\chi_0/\nu$, which is the constant~\footnote{Then perturbation theory shows that the first order correction $\alpha_1=0$ vanishes, and
hence, there is no contribution to $\cL_2$.},
\begin{align}
 \alpha_0 = \frac{\Omega/2}{\delta+i\kappa}.
\end{align} 
Using the annihilation and creation operators of the unmodulated oscillator, $b$ and $b^\dagger$, the interaction term of the Liouvillian reads
\begin{align}
 \cL_1 = \frac{1}{i\hbar}[F x,\rho].\label{eq:L1}
\end{align} 
In the last step we defined the operator
\begin{align}
 F= -\hbar g \left[a^\dagger a+\alpha_0 a^\dagger +
 \alpha_0^\ast a\right].
\end{align}  
acting on cavity states only. The further
calculations follow the standard theory of optomechanical laser
cooling~\cite{om:wilsonrae2008} and is summarized in App.~\ref{app:mecool}.

To proceed, we consider the slowly evolving subspace belonging to the eigenvalue $\lambda=0$, whose degeneracy is lifted in second order perturbation theory, hence determining the unique final state of laser cooling. The dynamics in this subspace selected by $\cP = \cP^{\lambda=0}_0$ is governed by the master equation
\begin{align}
 \cP\dot\rho(t) &= \sum_{\lambda\neq
 0}\cP\cL_1\cK_\lambda(t)\cL_1\cP\rho(t),
 \label{eq:effmeq}
\end{align}
with the superoperator
\begin{align}
 \cK_\lambda(t)=\int\limits_{0}^{\infty} d\tau\,
 e^{\lambda_\text{c}\tau}\left\{e^{\int\limits_{t-\tau}^{t}
 dt' \lambda_\text{m}(t')}\right\}\cP^\lambda_0.
 \label{eq:Klambda}
\end{align}
In the following we evaluate the latter expression for specific modulations of the mechanical frequency. 

\subsection{Periodic modulations} 
We assume that the frequency of the
mechanical oscillator $\nu(t)$ is changed periodically in time, that
is
\begin{align}
 \nu(t) = \nu_0 + f(t)
\end{align} with a real-valued periodic function $f(t+T) = f(t)=\sum_{l=-\infty}^\infty c_l e^{il\omega t}$, whereby $\omega
= 2\pi/T$. We further assume that the time averaged function $\langle
f(t)\rangle_t = 0$ vanishes. This can always be achieved if a
non-vanishing mean value is incorporated into $\nu_0$. Then, the term in the curly brackets in Eq.~\eqref{eq:Klambda} is also
periodic in $t$ with period $T$, and can be represented as a Fourier series
$\sum_\ell K_\ell^{\lambda_\text{m}}(t) \exp[i \ell\omega t]$. After performing the
$\tau$-integral, one finds
\begin{align}
 \cK_\lambda(t) = \sum_\ell K_\ell^{\lambda_\text{m}}(t)\frac{-\cP_0^\lambda}{\lambda_\text{c}+\lambda_\text{m}+i \ell\omega}.\label{eq:Klsin}
\end{align}

The evaluation of Eq.~\eqref{eq:effmeq} using the form~\eqref{eq:Klsin} again goes along the lines of laser cooling theory and is reported in App.~\ref{app:crates}. The resulting master equation of cooling contains the heating and cooling rates
\begin{align}
 A_\pm = \sum_\ell \langle K_\ell\rangle_t\frac{2\kappa\chi_0^2
 |\alpha_0|^2}{\left(\delta\mp[\nu_0+\ell\omega]\right)^2+\kappa^2},
 \label{eq:hcrates}
\end{align} 
which are already time-averaged over one period $T$ with
\begin{align}
 \langle K_l\rangle_t = \left(\frac{1}{2\pi}\right)^2\left|\int\limits_0^{2\pi} e^{-il\tau}\exp\left[{\sum_{m}\frac{c_m}{m\omega}e^{i m \tau}}\right]\,d\tau\right|^2.
\end{align}
The time averaging is justified when the modulation frequency belongs to the fast time scales of $\cL_0$, but still is slow enough to fulfill the adiabaticity condition.
Equation~\eqref{eq:hcrates} is the central result of this work. The rates of heating and cooling considered as a function of $\delta$ are a superposition of Lorentzians, centered at $\pm(\nu+\ell\omega)$ and weighted by the time-averaged Fourier-coefficients $\langle K_l\rangle_t$. In the resolved sideband-limit where $\kappa\ll\nu,\omega$, the Lorentzians are resolved and lead to several resonances in the cooling and heating of the mechanical element. In the case of sinusoidal modulation,
\begin{align}
 f(t) = \hat\nu \sin(\omega t)
\end{align}
the coefficients $K_l^{\lambda_\text{m}}(t)$ are up to a time-dependent phase given by Bessel functions
\begin{align}
 K_l^{\lambda_\text{m}}(t) = e^{-i \frac{k\hat\nu}{\omega}\cos\omega t}e^{-il\omega t}i^l J_l(k\hat\nu/\omega)
 \label{eq:ksint}
\end{align}
of order $l$ with the index $k$ determining the mechanical eigenvalue $\lambda_\text{m} = ik\nu_0$ of the unmodulated oscillator.
Time averaging leads to
\begin{align}
 \langle K_l\rangle_t = J_l(\hat\nu/\omega)^2\label{eq:klavg}
\end{align}
in the rates $A_\pm$.

\begin{figure}
  \centering
  \textbf{a)}
  \vtop{\vskip-1ex\hbox{\includegraphics[width=0.8\linewidth]{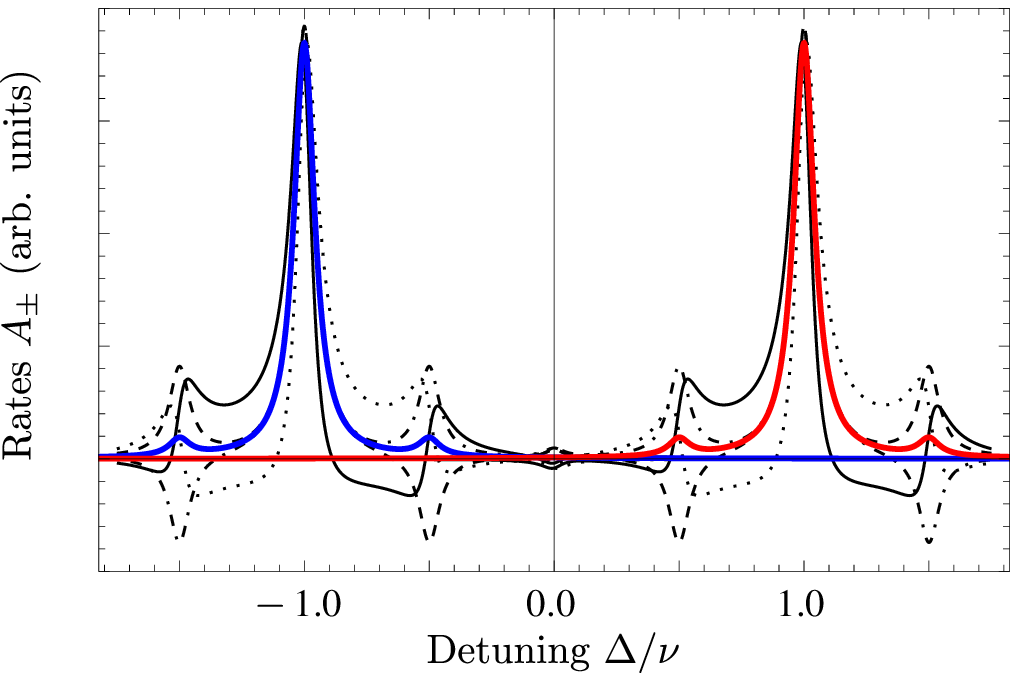}}}\\
  \textbf{b)}
  \vtop{\vskip-1ex\hbox{\includegraphics[width=0.8\linewidth]{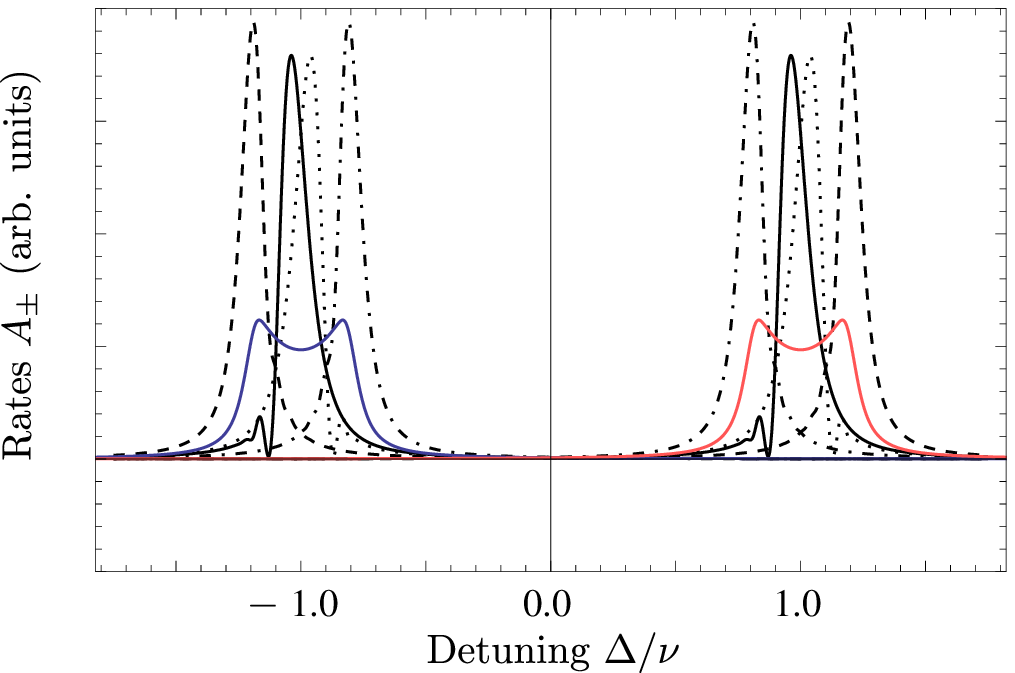}}}\\
   \caption{\label{fig:apm} The rates $A_\pm$ for a sinusoidal modulation. The blue (red) line represents the time average $A_-$ ($A_+$), Eq.~\eqref{eq:hcrates}. The black curves show the same rates before averaging at different times $t=0$, $T/4$, $T/2$ and $3T/4$ of a modulation period.  In \textbf{a)} the fast modulation $\omega=\nu/2$ generates rapidly oscillating sidebands displaced by multiples of $\omega$ from the main resonance at $\Delta = \pm\nu$. The sidebands have a non-vanishing temporal mean as shown by the colored curves. Slow modulations with $\omega=\nu/100$, in \textbf{b)}, essentially move the resonance peaks along the momentary oscillator frequency. Other parameters are $\hat \nu = 0.2\nu$, $\kappa=0.05\nu$.}
   
\end{figure}

Fig.~\ref{fig:apm} illustrates the dependency of the rates $A_\pm$, Eq.~\eqref{eq:hcrates}, on the detuning $\Delta$ (colored curves). Moreover, the rates before time-averaging, using Eq.~\eqref{eq:ksint}, for a $\sin$-modulation (black curves) are shown. In a) a fast modulation $\omega=\nu/2$ was chosen. The time-dependent curves show rapid oscillations of sideband peaks that even can become negative. The cooling dynamics can not follow these rapid oscillations, but after time-averaging positive sidebands at $\pm\omega$ around the main resonances at $\pm\nu$ survive, which influence the final state of laser cooling. On the contrary, for slow modulations $\omega = 0.01\nu$ presented in b), the sidebands overlap and lead to a periodically moving, positive resonance. In the resolved sideband-limit $\kappa\ll\nu$, the optimal detuning is given by the maximum of $A_-$, which sweeps through different values of $\Delta$.

\subsection{Stationary state}
\label{sec:stationarystate}
The time-averaged stationary state of the mirror's motion we calculate from the rates~\eqref{eq:hcrates} giving
\begin{align}
 \mu_{st} =
 \left(1-\frac{A_-}{A_+}\right)\left(\frac{A_-}{A_+}\right)^{b_0^\dagger
 b_0}
\end{align} 
with mean vibrational occupation number
\begin{align}
 \langle m\rangle = \Tr[b_0^\dagger b_0\mu_{st}] = \frac{A_+}{A_--A_+},\label{eq:mn}
\end{align}
which corresponds to the temperature at the end of the cooling procedure. The average cooling rate is given by
\begin{align}
 \Gamma = A_--A_+.\label{eq:GammaCool}
\end{align}
and estimates the time scale on which the stationary state is reached.

\begin{figure}
  \centering
  \includegraphics[width=\linewidth]{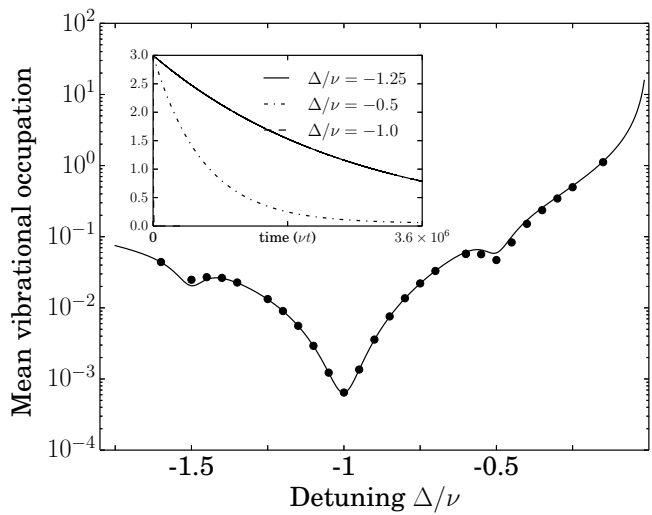}
  \caption{Comparison between the analytic approximation, Eq.~\eqref{eq:mn}, (solid
    line) and numerical results (points) for sinusoidal modulation. In the inset, the numerically calculated time curve $\langle m\rangle(t)$ is shown for selected values of $\Delta$. Parameters are
    $\hat{\nu}=0.1\nu$, $\kappa=0.05\nu$,
    $\omega=\nu/2$, $\gamma=0$, $\bar m=0$.}
  \label{fig:comparison}
\end{figure}

In Fig.~\ref{fig:comparison} we show the mean vibrational occupation number $\langle m\rangle$, Eq.~\eqref{eq:mn}, at the end of the cooling procedure (curve) and compare it with numerical solutions (dots) obtained from Eq.~\eqref{eq:mfromC}. Low phonon numbers are found whenever the detuning coincides with a resonance of $A_-$ in this resolved sideband limit. Small deviations are found at the sideband minima which we attribute to the adiabatic approximation performed in the theory. In the inset, the numerically calculated time dependency of the mean phonon number, starting from an initial state with $\langle m\rangle(t=0)=3$, is shown for different detunings. The decay rate of the curve corresponds to $\Gamma$, Eq.\eqref{eq:GammaCool}, and one finds indeed that the stationary state is approached faster close to a cooling resonance, where $A_-$ has a maximum.

Up to this point, the presented treatment ignored the thermalization of the mechanical oscillator due to coupling to its environment, which takes place with the rate $\gamma$. For high-Q oscillators, $\gamma$ can be considered as the slowest time scale involved in the dynamics. If this is the case, the final vibrational occupation number results as 
\begin{align}\label{eq:finaltemp}
 \overline{\langle m\rangle} = \frac{\langle m\rangle\Gamma_\text{cool}+\bar m \gamma}{\Gamma_\text{cool}+\gamma}.
\end{align}
In the adiabatic limit, $\Gamma_\text{cool}=\Gamma$, Eq.~\eqref{eq:GammaCool} can be calculated directly from $A_\pm$. 
 
 If the damping rate becomes comparable to the cooling rate, the damping should be incorporated into $\cL_1$, but, since $\cL_\gamma$ does not couple the $\cP$ and $1-\cP$ subspaces, the theory predicts no interplay between damping and optomechanical interactions. For even stronger damping, apart of the fact that the description in Eq.~\eqref{eq:Lgamma} becomes questionable, the laser cooling would take place on a much slower time scale, meaning that the mechanical element would take on a stationary state close the thermal state.

\subsection{Influence of pulse shape}

The strength of the sidebands in the rates $A_\pm$, Eq.~\eqref{eq:hcrates}, is determined by the coefficients $\langle K_l\rangle$ which in turn depend on the Fourier coefficients $c_m$ of the pulse $f(t)$ and can therefore be manipulated by altering the pulse form. By modifying the modulation parameters or using higher harmonics in $f(t)$ it is hence possible to shape the resonances of $A_\pm$. However, suppressing resonances in the heating rate $A_+$ for efficient cooling can not be achieved, since all $\langle K_l\rangle_t$ are non-negative.

\begin{figure}
  \centering
  \includegraphics[width=\linewidth]{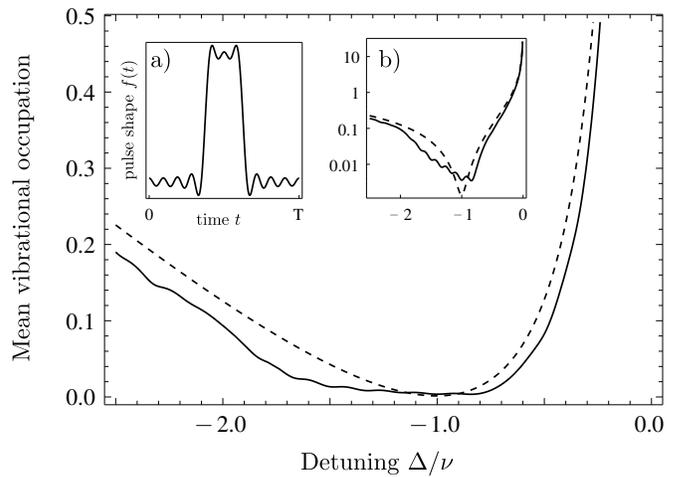}
   \caption{\label{fig:recpulse} Mean phonon number $\langle m\rangle$, Eq.~\eqref{eq:mn}, for a pulse $f(t)$, shown in inset \textbf{a)}, comprising several harmonics. It is compared to the unmodulated case (dashed curve). In inset \textbf{b)} a logarithmic plot shows that the modulation increases the minimal achievable temperature. The parameters are $\hat{\nu}=0.5\nu$, $\kappa=0.07\nu$, $\gamma=0$, $\bar m=0$, $\omega = \nu/6$.}
\end{figure}

Fig.~\ref{fig:recpulse} shows an example using the first ten Fourier coefficients of a rectangular shaped pulse, leading to a $f(t)$ as shown in inset a). The modulation frequency is chosen here to be smaller, $\omega = \nu /6$ such that with $\kappa=0.07\nu$ adjacent resonances in the rates $A_\pm$ start to overlap. In this way it is possible to broaden the cooling resonance and forming a plateau-like behavior of $\langle m\rangle(\Delta)$ around the lowest temperature. This can be achieved on the cost of an increased minimal temperature as the curves in the logarithmic scale of inset b) clearly shows.  

\section{Numerical results and discussion}
\label{sec:tierra}

In order to extend the analysis beyond the theoretical treatment, we
proceed with a numerical simulation of the optomechanical dynamics. We will analyze a large parameter regime of modulation
frequencies and incorporate the damping of the mechanical element.

\begin{figure}
\setlength{\unitlength}{1cm}
  \begin{picture}(8,13)
  \put(0,6.5){\includegraphics[width=8cm]{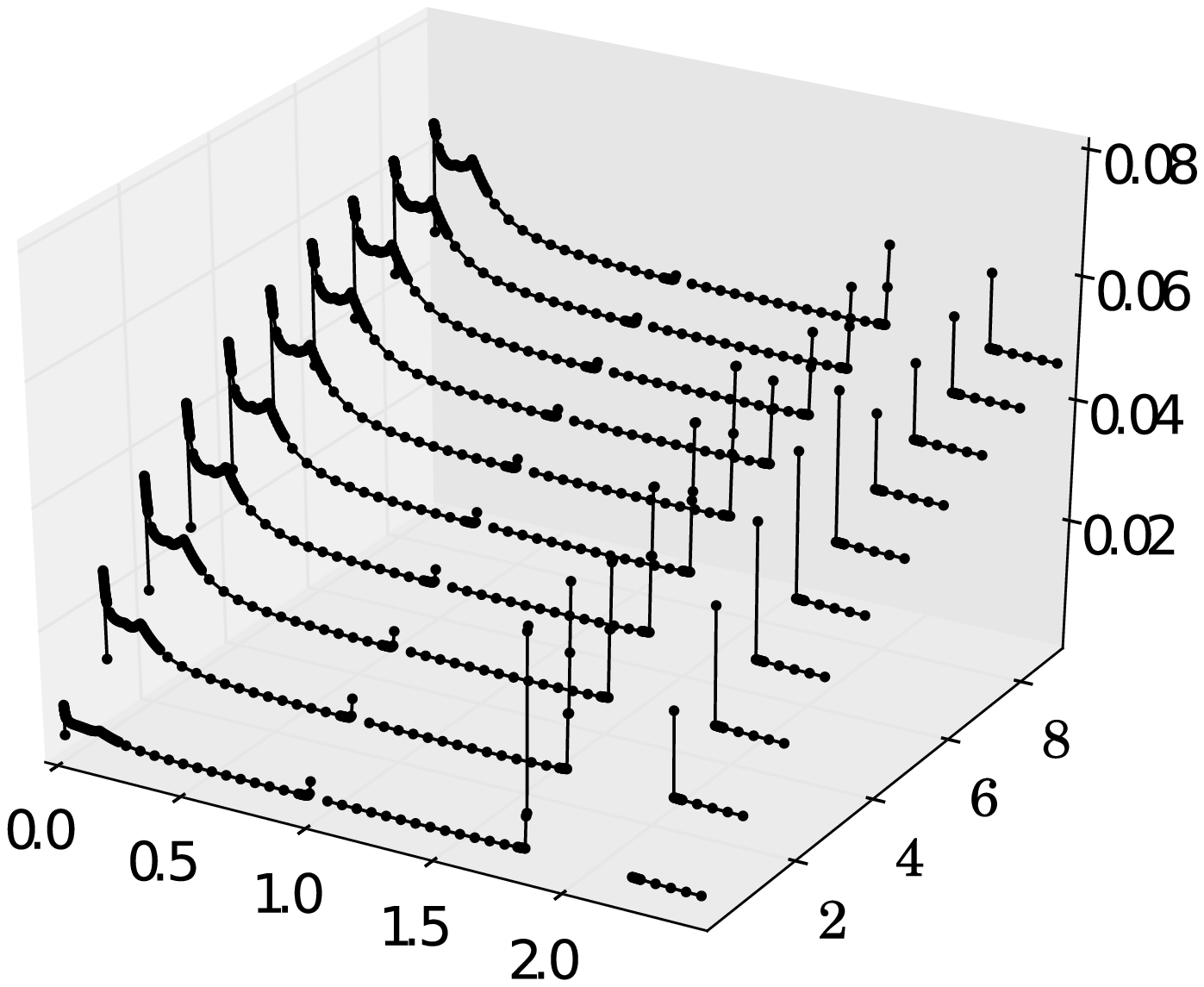}}
  \put(0,0){\includegraphics[width=8cm]{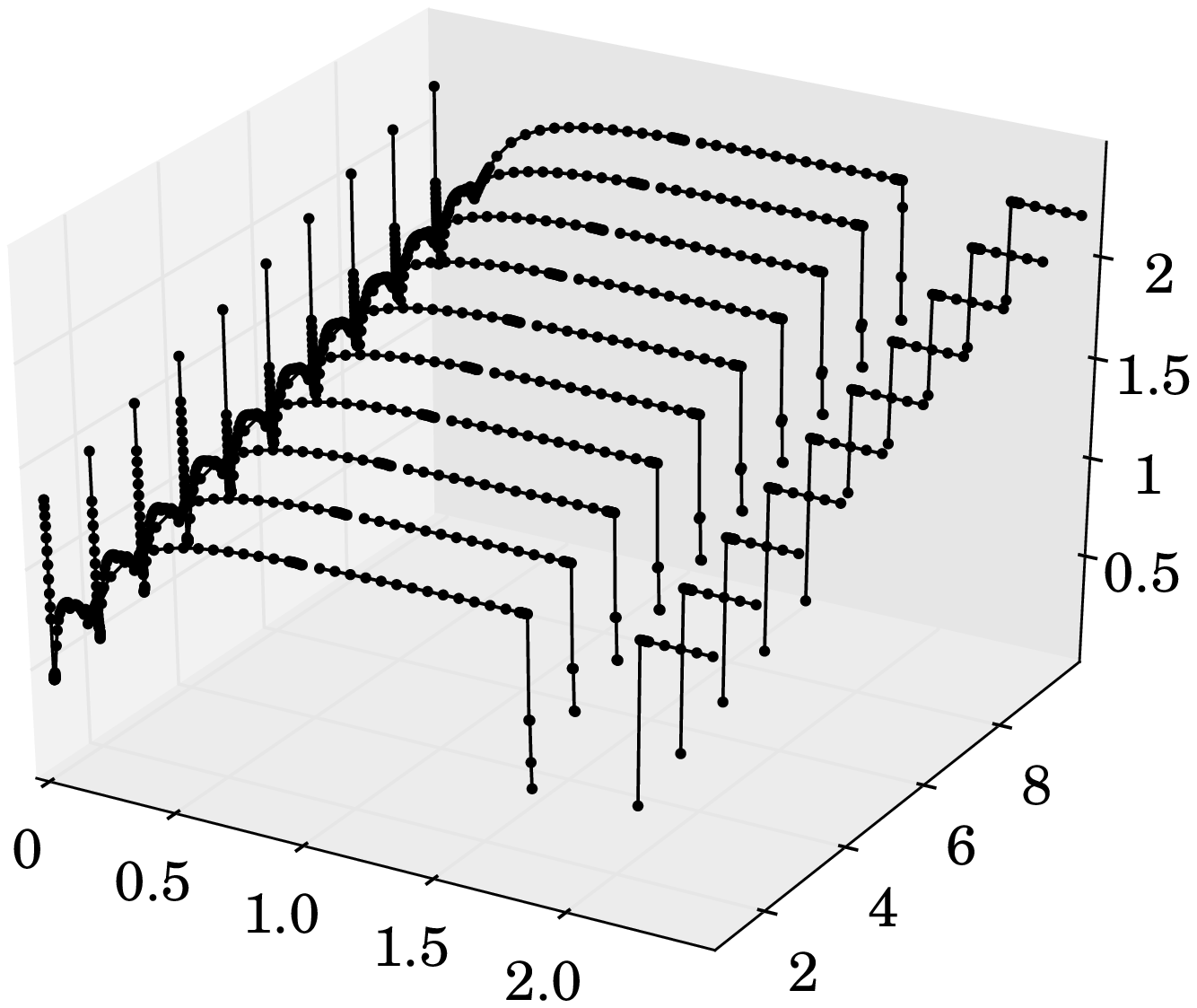}}
  \put(0,12){\textbf{a)}}
  \put(2,6.5){\large$\omega/\nu$}
  \put(6.15,7){\large$\gamma_0/\nu\times 10^{-3}$}
  \put(7.9,9.8){\rotatebox{90}{\large$\langle m\rangle$}}
  \put(0,5.5){\textbf{b)}}
  \put(7.8,1.8){\rotatebox{90}{\large$\Gamma_\text{cool}/\nu\times 10^{-2}$}}
  \put(2,0){\large$\omega/\nu$}
  \put(6.15,0.5){\large$\gamma_0/\nu\times 10^{-3}$}
  \end{picture}
  \caption{\textbf{a)} Mean vibrational occupations number as a function of the modulation frequency $\omega$ and damping $\gamma$. For each point, the detuning was optimized in order to find the minimal occupation number. Missing lines between the points mark regions where the system is instable. \textbf{b)} Same as in a) but showing the cooling rate $\Gamma_{\rm cool}$. Parameters:
    $\hat{\nu}=0.2\nu$, $\hat{\gamma}=0.2\gamma_0$, $\kappa=0.1\nu$ and $\eta
    = 0.25$ at resonance. }
  \label{fig:tierradesconocida}
\end{figure}

For the treatment discussed here, we focus again on a sinusoidal
modulation 
\begin{align}
  \label{eq:frecuency_modulation}
  \nu(t)&=\nu_0+\hat{\nu}\sin(\omega t)\, ,\\
  \gamma(t)&=\gamma_0+\hat{\gamma}\sin(\omega t)\, ,
\end{align}
and first solve numerically the coupled differential equations
\eqref{eq:dsysab}. By means of the behavior of the mean values
$\alpha(t)$ and $\beta(t)$ we are able to exclude parameter regimes
showing optomechanical instability~\cite{marquadt2006} or where
the system becomes unstable due to parametric amplification, as it is
found at the resonances $\omega=\nu$ and especially around
$\omega=2\nu$. The time-dependent mean values are used in the
linearized equation~\eqref{eq:dynC} in order to calculate the
time-averaged value of the mean vibrational occupation number $\langle
m\rangle$ in the quasi-stationary, oscillatory regime at the final
stage of the cooling. We also extract the cooling rate from the
numerical data by fitting an exponential decay to $\langle
m\rangle(t)$, calculated from Eq.~\eqref{eq:mfromC}.
For the rest of this section we
investigate the behavior of $\langle m\rangle$ and
$\Gamma_\text{cool}$ for a wide range of modulation frequencies and
dampings $\gamma_0$.
The results are depicted in Fig.~\ref{fig:tierradesconocida}a) where
we show the minimal value of the time-averaged occupation number
$\langle m\rangle$ after scanning through the detuning $\Delta$, as a
function of $\omega$ and $\gamma_0$ in the quasi-stationary regime.

For $\omega$ in the vicinity of zero, the vibrational occupation
number increases rapidly as $\omega$ increases. The change in the
oscillator frequency makes it impossible for the external field to
stay tuned to the optimal cooling frequency when $\kappa\ll \nu$,
explaining the sudden drop-off. This can be compared to the case
depicted in Fig.~\ref{fig:apm}b): The oscillatory moving resonance
curve (black lines) sweeps through the constant laser frequency such
that a fixed resonance condition is lost. When the modulation
frequency $\omega$ becomes larger, the time averaged resonance
curve,{\it i.e.} the colored curve in Fig.~\ref{fig:apm}b), becomes
relevant and the temperature decreases again, until it stays for a
small range of modulation frequency on a plateau-like level with
hardly visible small resonances, terminated by a final distinct
resonance peak at $\omega\approx0.15\nu$. Similar to the smaller
peaks, the final peak can be explained by the zeros of the Bessel
functions in the time-averaged rates $A_\pm$, Eqs.~\eqref{eq:hcrates}
and~\eqref{eq:klavg}. This becomes clear when recalling that the rates
$A_\pm$ can be decomposed into Lorentzian peaks with intensity
proportional to $J_l(\hat\nu/\omega)$, with $l=0$ marking the central
peak. For growing $\omega$ the argument of the Bessel function
decreases and from a certain modulation frequency on, the Bessel
function passed through its last zero and monotonically increases
afterwards. This modulation frequency marks the point from which on
the strength of the central peak only grows. When it overtops the
lateral peaks it uniquely defines the lowest temperature and this is
the point, where the final maximum of the plateau-like structure in
Fig.~\ref{fig:tierradesconocida}a) is situated. From here on, the
optimal detuning for lowest temperature is $\Delta=\nu$ and the
temperature falls off for larger values of $\omega$. Not before
$\omega=\nu$, another increase of the temperature can be observed in
the figure, due to a resonance with the vibrational frequency. At
$\omega=2\nu$ a strong increase of the temperature occurs. This is the
parameter regime discussed in detail in Ref.~\cite{farace:2012}, where
also squeezing of the oscillator's state and increased entanglement
between the optical and mechanical degree of freedom emerges. The
increase in temperature can qualitatively also be understood from the
rates $A_\pm$: Then, the first red sideband of the rate of heating
$A_+$ coincides with the central peak of the rate of cooling $A_-$.

Within the parameter range used here, no significant influence
stemming from the modulation of $\gamma(t)$ was found that could not
be reproduced by replacing $\gamma(t)$ with its average value
$\gamma_0$ for all considered time-averaged observables. 
The dependency of $\langle m\rangle$ on $\gamma_0$ mainly leads to an
overall increase of the vibrational occupation due to the stronger
coupling to the environmental heat bath. Apart from it, the course of
$\langle m\rangle$ for small $\omega$ shows a much stronger initial
increase of the temperature and the plateau-like behavior is
intensified, especially for mid-level values of $\gamma_0$. Also a
counterbalancing of the fluctuation of $\langle m\rangle$ around the
resonance at $\omega=\nu$ can be observed. Apart from these details, one
has to conclude that the damping of the modulated oscillator does not
show beneficial effects for cooling, but behaves as expected from a
stronger coupling to the heat bath.

In Fig.~\ref{fig:tierradesconocida}b) we plot the cooling rate
$\Gamma_{\rm cool}$ as a function of $\omega$ and $\gamma_0$. It shows
an overall reciprocal behavior compared to the final temperature shown
in~\ref{fig:tierradesconocida}b), as Eq.~\eqref{eq:GammaCool}
suggests. The cooling rate increases almost linearly with larger
$\gamma$, without changing significantly the shape with respect to
$\omega$. For very small modulations, the cooling rate falls off
strongly. Moreover, the resonance at $\omega=\nu$ is hardly visible,
but at $\omega=2\nu$ we find again that efficient cooling breaks down
in favor of increased non-classical properties of the optomechanical
system.

We finally remark that for larger values of $\kappa$ the fine
structure of the curves in Fig.~\ref{fig:tierradesconocida} are washed
out due to the stronger overlapping sidebands in $A_\pm$.

\section{Conclusions}
\label{sec:conclusions}

We studied in detail the effect of laser cooling when the frequency of
the harmonically supported mechanical element $\nu(t)$ and its
dissipation rate $\gamma(t)$ depend on time. We obtained analytic
approximations valid in the adiabatic limit when $\dot\nu(t)/\nu^2\ll
1$. When $\omega\approx\nu$ or larger and when the setup is in the
resolved sideband regime, several cooling resonances can be found,
whose relative strength depend on the pulse shape. The form of the
periodic modulation pulses allows to shape the cooling resonances as a
function of the laser frequency to a certain extent. With the help of
simple expressions for the time-averaged rates of heating and cooling,
a physical picture for the temporal behavior of the cooling process
could be provided.

We complemented the analysis of cooling a modulated optomechanical
setup by numerical investigations covering a larger range of
parameters, whose results generally confirm the analytic predictions.
At least qualitatively, the insight provided by the theoretical model
allows to explain most of the numerical findings. Moreover, we found
that the time dependency of $\gamma(t)$ does not alter qualitatively
neither the final temperature nor the decay rates of the cooling
process. If the interest is only in the time-averaged behavior, the
modulation of the damping can even be ignored. The analysis of this
work provides a deeper insight into the cooling dynamics when the
properties of the mechanical oscillator are modulated and can serve as
a guidance for current experiments when the temporal behavior of the
mechanical element are controllable in time.

\begin{acknowledgments} We thank Phillipp Treutlein and Andreas
  J\"ockel for drawing our attention on this topic. Moreover, we are
  thankful to Luis Octavio Casta\~nos for many helpful discussions.
  This work was supported by project PAPIIT IN103714 and was triggered
  during the the Kolleg ``Open quantum systems: Chaos and
  decoherence'' at the Centro Internacional de Ciencias, Cuernavaca, Mexico, funded by the Alexander-von-Humboldt
  foundation.

\end{acknowledgments}

\begin{appendix}

\section{Elements of the linearized master equations}
\label{app:linme}
The explicit forms of the matrices and vectors appearing in Eqs.~\eqref{eq:meqf} follow directly from the master equation~\eqref{eq:me} in the linearized form, {\it i.e.}, after omitting the term proportional to $a^\dagger a$ in the optomechanical interaction. One finds 
\begin{align}
 \hat H = \hbar\begin{pmatrix}
           -\frac{\Delta}{2} & 0 & -\frac{\chi_0}{2}(\alpha+\alpha^\ast) & 0 \\
           0 & -\frac{\Delta}{2} & -\frac{i\chi_0}{2}(\alpha^\ast-\alpha) & 0 \\
           -\frac{\chi_0}{2}(\alpha+\alpha^\ast) & -\frac{i\chi_0}{2}(\alpha^\ast-\alpha) & \frac{\nu}2 & 0 \\
           0 & 0 & 0 & \frac{\nu}{2}
          \end{pmatrix}
\end{align}
and for the jump vectors
\begin{align}
 \vec L_1 = \frac{1}{\sqrt 2}\begin{pmatrix} 1 \\ i \\ 0 \\ 0 \end{pmatrix} ,\quad
 \vec L_2 = \frac{1}{\sqrt 2}\begin{pmatrix} 0 \\ 0 \\ 1 \\ i \end{pmatrix} ,\quad
 \vec L_3 = \frac{1}{\sqrt 2}\begin{pmatrix} 0 \\ 0 \\ 1 \\ -i \end{pmatrix}.
\end{align}
Furthermore, we defined
\begin{align}
 \gamma_1 = 2\kappa,\quad
 \gamma_2 = \gamma(\bar m + 1),\quad
 \gamma_3 = \gamma \bar m.
\end{align}
The matrix $(\hat\Gamma)_{mn}=\sum_k \gamma_k (\vec L_k)^\ast_m (\vec L_k)_n$ is needed for the time evolution, Eq.~\eqref{eq:dynC}, of the covariance matrix $\hat C$. It has the explicit form
\begin{align}
 \Gamma = \frac 12\begin{pmatrix}
          \kappa & i\kappa & 0 & 0 \\
          -i\kappa & \kappa & 0 & 0 \\
          0 & 0 & \frac{\gamma}2(2\bar m+1) & \frac{i}{2}\gamma\\
          0 & 0 & -\frac{i}{2}\gamma & \frac{\gamma}2(2\bar m+1)
          \end{pmatrix}.
\end{align}
With these quantities the time evolution of Gaussian states can be written in the compact notation given in Sec.~\ref{sec:linint}.

\section{Derivation of effective master equation}
\subsection{Spectral decomposition of $\cL_0$}
\label{app:specdecomp}

In this section we discuss the eigensystem of $\cL_0=\cL_\text{c}+\cL_\text{m}(t)$, Eq.~\eqref{eq:L0}. To this end we introduce the eigenelements $\hat\rho^{\lambda}(t)$ and $\check\rho^{\lambda}(t)$ which are elements of the operator space and fulfill instantenous the eigenvalue equations 
\begin{align}
 \cL_0(t) \hat\rho^{\lambda}(t) &= \lambda(t)\hat\rho^{\lambda}(t)\\
 \check\rho^{\lambda\dagger} \cL_{0}(t) &=
 \lambda(t)\check\rho^{\lambda\dagger}(t).
\end{align}
These eigenelements are orthogonal
$\Tr\{(\check\rho^{\lambda})^\dagger\hat\rho^{\lambda'}\}=\delta_{\lambda,\lambda'}$
with respect to the scalar product $(A,B) = \Tr[A^\dagger B]$. The projectors 
$\cP_0^\lambda=\check\rho^\lambda\otimes\hat\rho^\lambda$ project an arbitrary operator $X$ onto the subspace belonging to $\lambda$ according to $\cP_0^\lambda X = \hat \rho^\lambda \Tr[\check \rho^{\lambda\dagger}X]$. If the eigensystem of $\cL_0$ is complete, the projectors fulfill
\begin{align}
 \sum_\lambda \cP^\lambda_0 = 1.
\end{align} 
Since in lowest order perturbation theory $\cL_0$ is a sum of the cavity and mechanical part without interaction, the eigensystem can be decomposed into
\begin{align}
 \hat\rho^{\lambda}(t)  &=  \hat\rho^{\lambda}_\text{c}  \hat\rho^{\lambda}_\text{m}(t),\quad
 \check\rho^{\lambda\dagger}(t)  =  \hat\rho^{\lambda\dagger}_\text{c}  \hat\rho^{\lambda\dagger}_\text{m}(t),\\
 \cP_0^\lambda(t) &=  \cP_\text{c}^\lambda \cP_\text{m}^\lambda(t).
\end{align}
and
\begin{align}
 \lambda_0(t) = \lambda_\text{c} + \lambda_\text{m}(t).\label{eq:lambda0app}
\end{align}
The explicit form of the cavity eigenelements can be found
in Ref.~\cite{qo:briegel1993} with the corresponding eigenvalues
\begin{align}
 \lambda_\text{c}(k,n) = ik\Delta -(2 n+|k|)\kappa,
\end{align}
where $k$ are integers and $n$ non-negative integers. The mechanical eigenelements can be constructed from the
instantaneous eigenstates $\ket{m(t)}$ of $H_\text{mec}(t)$ and have the simple form
\begin{align}
  \hat\rho^{\lambda}_\text{m}(t; n,l) = 
  \begin{cases}
   \ket{n}\bra{n+l} & l\ge 0\\
   \ket{n+|l|}\bra{n} & l<0
  \end{cases}
\end{align}
with $\lambda_\text{m}(t; n,l) = i l \nu(t)$ and $\check\rho_\text{m}^{\lambda_{\rm m}}=\hat\rho_\text{m}^{\lambda_{\rm
m}}$. These eigenelements are infinitely degenerate in the index $n$.

\subsection{Effective master equation of cooling}
\label{app:mecool}
We start with the
 Liouvillian from Eq.~\eqref{eq:cl01} and define the time-independent
 projectors
 \begin{align}
  \cP &=
  \cP_\text{c}^{\lambda_\text{c}=0}\cP_\text{m}^{\lambda_\text{m}=0}\\
  \cQ &= 1-\cP
 \end{align} in the adiabatic approximation discussed in the main
 text. The goal is to find an effective master equation in the
 subspace selected by $\cP$ valid up to second order in the parameter
 $\eta$ characterizing the optomechanical coupling strength. We
 project the master equation $\dot\rho=\cL\rho$ onto the subspaces
 belonging to $\cP$ and $\cQ$, insert $1=\cP+\cQ$, and use the
 expansion~\eqref{eq:cl01} of $\cL$, yielding
 \begin{align}
  \cP\dot\rho(t) &= \cP\cL_1\cQ\rho(t)\label{eq:Peq}\\
  \cQ\dot\rho(t) &= \cQ\cL_1\cP\rho(t) + \cQ\cL_0\cQ\rho(t).
 \end{align} In the last step we exploited that $\cP\cL_0=\cL_0\cP =
 0$ with the aproximated $\cL_0$ from Eq.~\eqref{eq:adiabapp} and that $\cL_1$ does not couple states in $\cP$, hence
 $\cP\cL_1\cP = 0$. Moreover, we neglected $\cL_1$ in the last term of
 the second equation, since it would lead to higher order correction.
 The second equation can be formally integrated leading to
\begin{align}
 \cQ\rho(t) = &e^{\int\limits_{t_0}^t dt' \cQ\cL_0(t')}
 Q\rho(t_0)\nonumber\\ +&e^{\int\limits_{t_0}^t dt'
 \cQ\cL_0(t')}\int\limits_{t_0}^{t} dt'e^{-\int\limits_{t_0}^{t'} dt''
 \cQ\cL_0(t'')}\cQ\cL_1P\rho(t').
\end{align} The integration interval $\Delta t = t-t_0$ is assumed
here to be large for the fast time scale of zeroth order and short for
the time scale of interaction: The first term thus dies off rapidly
whereas in the second term one can replace $P\rho(t')$ by $P\rho(t)$.
Plugging this result into Eq.~\eqref{eq:Peq} and inserting the
completeness relation $\sum_\lambda \cP^\lambda=1$ yields the closed
master equation of the main text, Eq.~\eqref{eq:effmeq},
where we further used Eq.~\eqref{eq:lambda0app} and set $\Delta t\to \infty$ in
the limits of the $\tau$-integral.

\subsection{Cooling and heating rates}
\label{app:crates}

Inserting Eq.~\eqref{eq:Klsin} into the effective master equation~\eqref{eq:effmeq} and tracing over the cavity degrees of freedom yields
\begin{align}
 \dot\mu = \sum_{\lambda\neq 0} \sum_l \frac{K_l^{\lambda_\text{m}}}{\lambda_\text{c} + \lambda_\text{m}+i l \omega}\Tr_c\{\cP\cL_1\cP_0^\lambda\cL_1\varrho_\text{c}\mu\}\label{app:mudot}
\end{align}
with the density operator $\mu=\Tr_\text{c} \cP\rho$ and the stationary state $\varrho_\text{c}=\ketbra{0}{0}$ of the cavity, full filling $\cL_\text{c}\varrho_\text{c}=0$. The coefficients fulfill $K_l^{\lambda_\text{m}}(t)=K_{-l}^{-\lambda_\text{m}\ast}(t)$. The trace is calculated using Eq.~\eqref{eq:L1}, and after separating operators belonging to the oscillator and cavity degrees of freedom, the trace term reads
\begin{align}
 \Tr_\text{c}\{\dots\} = T_1(\lambda_\text{c})\cP_m^0 [x,\cP_m^{\lambda_\text{m}} x\mu]- T_2(\lambda_\text{c})\cP_m^0[x,\cP_m^{\lambda_\text{m}} \mu x],
\end{align}
whereby
\begin{align}
 T_1(\lambda_\text{c}) &= \Tr_\text{c}\{F \cP_\text{c}^{\lambda_\text{c}}F\varrho_\text{st}\}=\hbar^2 g^2 |\alpha_0|^2\delta_{\lambda_\text{c},i\delta-\kappa}, \label{eq:appT1}\\
 T_2(\lambda_\text{c}) &= \Tr_\text{c}\{F \cP_\text{c}^{\lambda_\text{c}}\varrho_\text{st}F\} =\hbar^2 g^2 |\alpha_0|^2\delta_{\lambda_\text{c},-i\delta-\kappa}.
\end{align}
The mechanical expressions
\begin{align}
 \cP_m^0 [x,\cP_m^{\lambda_\text{m}} x\mu] &= \xi_0^2\Big\{(b_0b_0^\dagger -b_0^\dagger \mu b_0)\delta_{\lambda_\text{m},-i\nu}\nonumber\\
 &\qquad+(b_0^\dagger b_0\mu-b_0\mu b_0^\dagger) \delta_{\lambda_\text{m},i\nu}\Big\},\\
 \cP_m^0 [x,\cP_m^{\lambda_\text{m}} \mu x] &= \xi_0^2\Big\{(b_0\mu b_0^\dagger -\mu b_0^\dagger b_0)\delta_{\lambda_\text{m},-i\nu}\nonumber\\
 &\qquad+(b_0^\dagger \mu b_0-\mu b_0 b_0^\dagger) \delta_{\lambda_\text{m},i\nu}\Big\}\label{eq:appPm}
\end{align}
are calculated with $x=\xi_0(b_0+b_0^\dagger)$. After plugging Eqs.~\eqref{eq:appT1}--\eqref{eq:appPm} into Eq.~\eqref{app:mudot}, performing the time averaging with $\langle K_l\rangle_t \equiv\langle K_l^{\lambda_\text{m}=i\nu}\rangle = \langle K_{-l}^{\lambda_\text{m}=-i\nu}\rangle$, the master equation can be cast into the form
\begin{align}
 \dot\mu = \frac{1}{i\hbar}[\widetilde{H},\mu]+\frac{A_-}2\cD[b_0]\mu+\frac{A_+}2\cD[b_0^\dagger]\mu
\end{align}
of a damped harmonic oscillator with the rates $A_\pm$ given by Eq.~\eqref{eq:hcrates} and a small correction $\widetilde{H}\propto b_0^\dagger b_0$ to the Hamiltonian of the mechanical oscillator, that is neglected in what follows.

\end{appendix}

\end{document}